\documentclass[10pt, journal,a4paper]{IEEEtran}
\usepackage{amsmath}
\usepackage{amssymb}
\usepackage{cite}
\usepackage{graphicx, subfigure}
\usepackage{psfrag}
\usepackage{url}
\usepackage[latin1]{inputenc}
\usepackage[absolute,overlay]{textpos}
\usepackage{textcomp}
\usepackage{gensymb}
\usepackage{hyperref}
\usepackage{bm}

\usepackage{pgf}

\usepackage[nocomma]{optidef}

\usepackage{algorithm}
\usepackage{algorithmic}

\makeatletter
\def\thickhline{%
  \noalign{\ifnum0=`}\fi\hrule \@height \thickarrayrulewidth \futurelet
   \reserved@a\@xthickhline}
\def\@xthickhline{\ifx\reserved@a\thickhline
               \vskip\doublerulesep
               \vskip-\thickarrayrulewidth
             \fi
      \ifnum0=`{\fi}}
\makeatother

\newlength{\thickarrayrulewidth}
\usepackage{amsthm}

\begin{document}
    \title{Movable Antennas-aided Wireless Energy Transfer for the Internet of Things}
\author{
	\IEEEauthorblockN{Osmel Mart\'{i}nez Rosabal, Onel Alcaraz L\'{o}pez, Marco Di Renzo, Richard Demo Souza, and Hirley Alves}
	\thanks{Osmel Mart\'{i}nez Rosabal, Onel Alcaraz L\'{o}pez, and Hirley Alves are with the Centre for Wireless Communications (CWC), University of Oulu, Finland. (email: firstname.lastname@oulu.fi)} 
        \thanks{M. Di Renzo is with Universit\'e Paris-Saclay, CNRS, CentraleSup\'elec, Laboratoire des Signaux et Syst\`emes, 3 Rue Joliot-Curie, 91192 Gif-sur-Yvette, France. (email: marco.di-renzo@universite-paris-saclay.fr), and with King's College London, Centre for Telecommunications Research -- Department of Engineering, WC2R 2LS London, United Kingdom (email: marco.di\_renzo@kcl.ac.uk)}
        \thanks{R. D. Souza is with the Department of Electrical and Electronics Engineering, Federal University of Santa Catarina (UFSC), Florian\'{o}polis, SC, Brazil. (e-mail: richard.demo@ufsc.br).}
	\thanks{This work is partially supported by Research Council of Finland (Grants 348515 and  369116 (6GFlagship)), by RNP/MCTI Brasil 6G project (01245.020548/2021-07), and by the Nokia Foundation, the French Institute of Finland, and the French Embassy in Finland under the France-Nokia Chair of Excellence in IC. The authors also wish to acknowledge CSC - IT Center for Science, Finland, for computational resources.}
}   
\maketitle

\begin{abstract}
Recent advancements in movable antennas (MAs) technology create new opportunities for 6G and beyond wireless systems. MAs are promising for radio frequency wireless energy transfer because they can dynamically adjust antenna positions, improving energy efficiency and scalability. This work aims to minimize the power consumed by an analog beamforming power beacon equipped with independently-controlled MAs (IMAs) for charging multiple single-antenna devices. To this end, we enforce a minimum separation among antennas and a minimum received power at the devices. The resulting optimization problem is nonlinear and nonconvex due to interdependencies among the variables. To tackle this, we propose a semidefinite program guided particle swarm optimization (SgPSO) algorithm where each particle represents an antenna configuration, and the fitness function optimizes the corresponding power allocation. SgPSO is utilized for configuring the MAs largely outperforming fixed array implementations, particularly with more antennas or devices. We also present an alternative implementation using uniformly-spaced MAs, whose performance closely approaches that of the IMAs, with the gap widening only as the number of devices grows. We also examine how increasing the number of antennas promotes near-field conditions, which decrease as devices become more widely distributed.

\end{abstract}

\vspace{-0.9em}
\section{Introduction}
Radio-frequency (RF) wireless energy transfer (WET) is a pivotal technology for Internet of Things (IoT) networks \cite{Lopez.2023} by eliminating the need for frequent maintenance, supporting a deploy-and-forget paradigm. 
Despite its potential, coverage remains a challenge due to its relatively low conversion efficiency and the effective transmit power regulations to avoid interference or potential harm \textit{in living species} \cite{international2020guidelines}. 

From the transmitter perspective, movable antennas (MAs) hold the potential to enhance energy efficiency with modest complexity. Unlike conventional arrays with fixed antennas, MAs can reconfigure their geometrical and electromagnetic characteristics, including the operating frequency and radiation patterns, via software. MAs enable transmitters to ``escape" deep fades and non-line-of-sight conditions \cite{Khammassi.2023}, achieving a similar performance with fewer antennas or power-consuming RF chains \cite{Zhang.2025,Ma.2024}, and in some cases, even with a reduced antenna aperture at the transmitter \cite{Zhu2.2024} compared to traditional arrays with non-reconfigurable antennas. These advantages have inspired flexible reflecting surfaces \cite{Mursia.2025} and antenna arrays \cite{Jiancheng.2025} whose shape can be modified in a three-dimensional space to dynamically control electromagnetic waves.

The applications of MAs span various domains, including industrial IoT, smart homes, robotic networks, satellite communications \cite{Zhu.2024}, integrated sensing and communications \cite{Ye.2025}, and over-the-air computation \cite{Zhang2.2024}. Recently, MA technology has been also proposed for RF-WET \cite{Gao.2024,Zhou.2024,Chen.2024}. For example, \cite{Gao.2024} investigates a wireless-powered communication network, wherein the antennas' positions and time allocation for energy/information transmission are configured to maximize the network's achievable throughput. Interestingly, the performance of a discrete step size MAs approaches its continuous counterpart when the stepper motor resolution and/or the array size increase. Moreover, \cite{Zhou.2024} studies an MA-aided simultaneous wireless information and energy transfer system aiming to maximize the receiver's throughput while satisfying a target power requirement. Results highlight the advantages of equipping all nodes with MAs, especially the transmitters. Finally, \cite{Chen.2024} investigates a mobile edge computing system with an MA charging multiple devices that offload individual tasks to the edge server. Results show that having a common configuration for RF-WET and task offloading is less effective than employing dedicated configurations for each service.

In this work, we study an independently-controlled MAs (IMAs)---in which each element move independently from the others---aided RF-WET system to charge a network of single-antenna IoT devices. We consider an analog beamforming architecture at the RF transmitter, referred to as a power beacon (PB) hereinafter, in which a single RF chain feeds the energy-carrying signal to a set of antennas disposed on a plane. We aim to minimize the PB's power consumption by jointly optimizing the antennas' positions and the transmit analog precoder vector subject to a minimum device's received power requirement. Unlike previous works, \textit{e.g.}, \cite{Gao.2024,Zhou.2024,Chen.2024}, this work focuses on the operation of the RF-WET service while considering the impact of the network deployment on the channel conditions. Our contributions are five-fold: i) we propose an analog beamforming MA-equipped PB architecture and formulate a nonconvex power consumption minimization problem accounting for the signal transmission phase; ii) we propose a semidefinite program (SDP) guided particle swarm optimization (PSO) algorithm, termed SgPSO, to tackle the coupling among optimization variables; iii) we propose uniformly-spaced MAs (UMAs) as an alternative to reduce hardware/configuration complexity of that of IMAs; iv) we show that the power consumption of the PB with MAs outperform benchmark architectures equipped with uniform arrays of fixed antennas, while the performance gap between IMAs and UMAs remains small and only widens as the number of served devices increases; and finally vi) we show how increasing the number of antennas promotes near-field operation, whereas expanding the service area and thus dispersing devices may reduce the probability of near-field conditions for the MAs.

\section{System Layout and Problem Formulation}

%
\begin{figure}
    \centering
    \includegraphics[width=\linewidth]{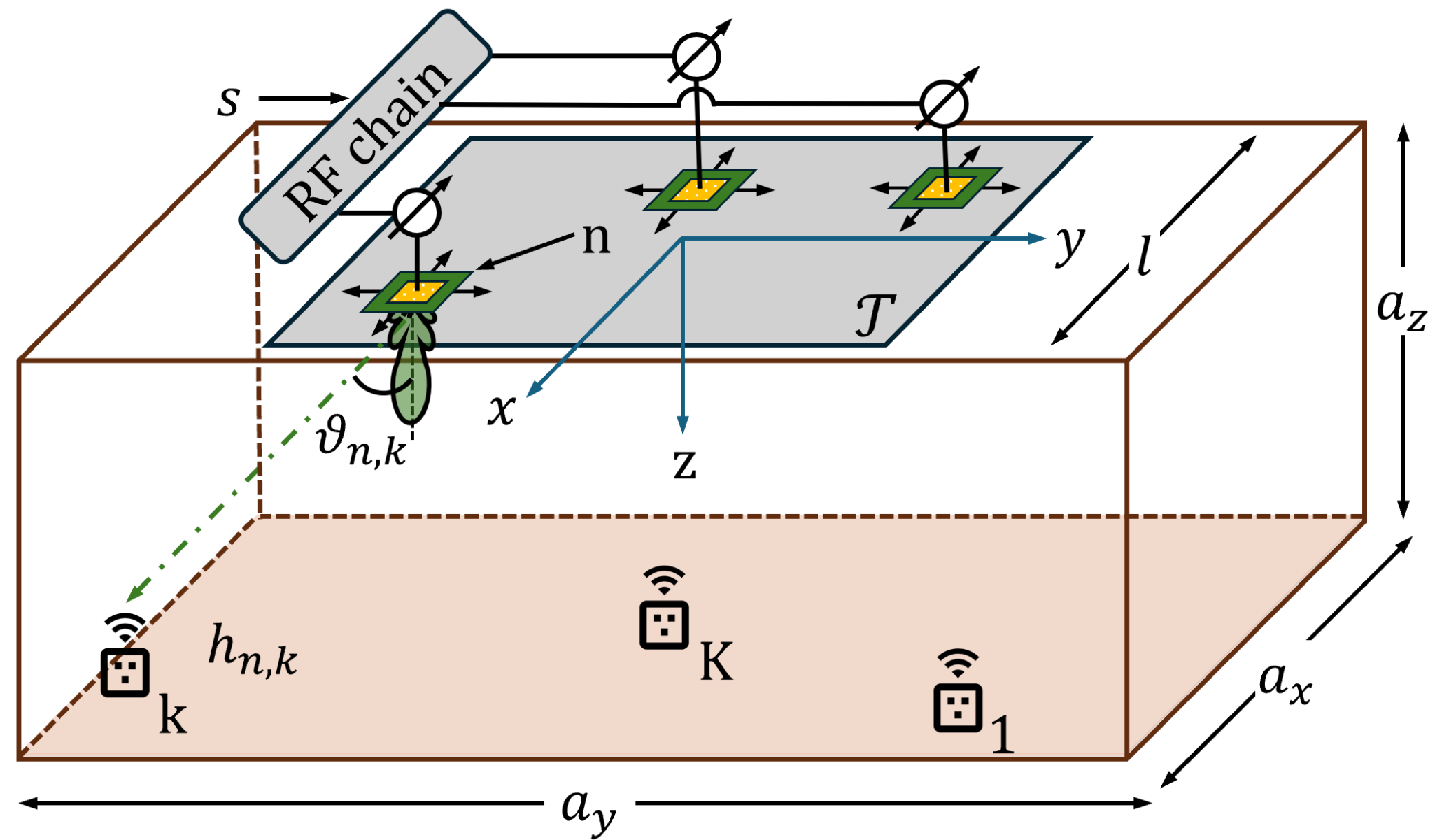}
    \vspace{-1em}
    \caption{PB equipped with IMAs charging a network of IoT devices.}
    \label{fig:systemModel}
\end{figure}
Consider the RF-WET system illustrated in Fig.~\ref{fig:systemModel}, where a PB equipped with a set $\mathcal{N}$ of $N$ IMAs concurrently charges a set $\mathcal{K}$ of $K$ single-antenna IoT devices. This configuration could represent an industrial scenario with the PB installed on the ceiling of a manufacturing facility to charge a network of sensors and monitor different stages of a production process. A single RF chain feeds the IMAs and can move within the square region $\mathcal{T}$ contained by the $x-y$ plane, centered at the origin, with side length $l$. We assume that the antennas move with an arbitrary small step size, allowing ideal system performance exploration, \textit{i.e.}, no hardware limitations. Herein, $\mathbf{r}_n \in \mathbb{R}^{2 \times 1}$ and $\mathbf{u}_k \in \mathbb{R}^{3 \times 1}$ denote the Cartesian coordinates of the $n^\mathrm{th}$ PB's antenna and the $k^\mathrm{th}$ device, respectively, whereas $\mathbf{h}_k \in \mathbb{C}^{N \times 1}$ captures the channel coefficients between the PB's antennas and the $k^\mathrm{th}$ device. Considering that the PB transmits a signal $s$ with power $p_T$ weighted by the unit-power analog precoder $\mathbf{w} = \frac{1}{\sqrt{N}}e^{j\bm{\theta}} \in \mathbb{C}^{N \times 1}$, where $\bm{\theta} \in \mathbb{R}^{N \times 1}$ is the phase shift vector, and ignoring the negligible noise power, the received power at the $k^\mathrm{th}$ device is
\begin{equation}
    p_k = \mathbb{E}_{s} \left[|\mathbf{h}^H_k \mathbf{w} s|^2\right] = p_T|\mathbf{h}^H_k \mathbf{w}|^2.
\end{equation}
The $n^\mathrm{th}$ element of $\mathbf{h}_k$ is modeled as \cite{Zhang.2022}
\begin{equation}
    h_{k,n} = \sqrt{F(\vartheta_{n,k},\kappa)}\frac{\lambda \exp\left(-\frac{2\pi j}{\lambda} \left\lVert \begin{bmatrix} \mathbf{r}_n \\ 0 \end{bmatrix} - \mathbf{u}_k \right\rVert_2\right)}{4\pi\left\lVert \begin{bmatrix} \mathbf{r}_n \\ 0 \end{bmatrix} - \mathbf{u}_k \right\lVert_2}, \label{eq:channelModel}
\end{equation}
where
\begin{equation}
    F(\vartheta_{n,k},\kappa) = \begin{cases}
        2(\kappa+1)\cos^\kappa \vartheta_{n,k}, \ &\vartheta_{n,k} \in \left[0,\frac{\pi}{2}\right] \\
        0, \ &\text{otherwise}
        \end{cases}
\end{equation}
is the radiation profile of the antennas. Herein, $\vartheta_{n,k}$ is the polar angle of the $k^\mathrm{th}$ device relative to the $n^\mathrm{th}$ antenna, $\kappa \geq 2$ is the boresight gain of the antennas, and $\lambda$ is the wavelength of the transmitted signal. Herein, we assume that the devices' positions are known (or can be acquired) at the PB, which allows determining the channel coefficients using \eqref{eq:channelModel}. 

\subsection{Problem Formulation}\label{subSec:problemFormulation}
We aim to configure the antenna positions and precoding to minimize the PB's power consumption subject to the receive power requirements $\{p_k^\mathrm{th}\}$ of the devices. We note that the antennas' positions remain fixed during the charging service unless the number of devices changes, their positions, and/or power requirements trigger the optimization procedure. The optimization problem can be formulated as follows:
\begin{subequations}\label{P1}
    \begin{alignat}{2}
    \mathbf{P1:} &\underset{\mathbf{w},\{\mathbf{r}_n\},p_T}{\mathrm{min.}} \ && p_T \label{P1a}\\
    &\text{s.t.} \ && p_k \geq p_k^\mathrm{th} , \ \forall k \in \cal K, \label{P1b}\\
    & \ && |w_n|^2 \!=\! \frac{1}{N}, \ n \in \mathcal{N}, \label{P1c}\\
    & \ && \lVert \mathbf{r}_n - \mathbf{r}_{n'} \rVert_2 \geq \delta, \ n, n' \in \mathcal{N}, \ n \neq n', \label{P1d} \\
    & \ && \mathbf{r}_n \in \mathcal{T}, \ \forall n \in \mathcal{N}, \label{P1e}
    \end{alignat}
\end{subequations}
where \eqref{P1c} is the constant modulus constraint for the analog precoder, \eqref{P1d} ensures a minimum separation $\delta$ among antennas to avoid coupling, and $\mathcal{T}$ denotes the set of feasible antenna positions, which is assumed to be convex. Notice that $\mathbf{P1}$ is nonconvex due to the nature of constraints $\eqref{P1b}-\eqref{P1d}$. 

The nonlinearity of \eqref{P1b} hinders finding the optimal solution of $\mathbf{P1}$ since these constraints are tightly coupled to the antenna positions and the analog precoder variables. Solving $\mathbf{P1}$ using conventional alternating optimization yields poor results, as the partial solutions from one subproblem can restrict the solution space of subsequent optimization steps \cite{Xiao.2024}.

\section{SDP-guided PSO}\label{sec:PSOSolution}
We propose a PSO-based solution that exploits its intrinsic exploration/exploitation features to optimize the antenna positions. Unlike standard PSO, which works over all variables simultaneously, our approach embeds a dedicated power allocation step to optimize the precoder for each candidate antenna constellation a particle represents. The steps described below are summarized in Algorithm~\ref{alg:PSO}, referred to as SgPSO. 

Let $\mathcal{Q} = \{\mathbf{Q}_1,\ \ldots,\ \mathbf{Q}_m,\ \ldots,\ \mathbf{Q}_M\}$ denote the set of particles where $\mathbf{Q}_m \in \mathbb{R}^{2 \times N}$ is a candidate antenna position, \textit{i.e.}, $\mathbf{Q}_m = [\mathbf{r}_{1,m},\ \mathbf{r}_{2,m},\ \ldots, \ \mathbf{r}_{N,m}]$. Each particle is associated with a velocity matrix $\mathbf{V}_m \in \mathbb{R}^{2 \times N}$ to change the particle's position in the quest to find the optimal solution. 

Moving from randomly distributed particles belonging to the set $\mathcal{T}$, the SgPSO algorithm keeps track of the particles' best $\{\mathbf{Q}_{\mathrm{pbest},m}\}$ and the global best $\mathbf{Q}_\mathrm{gbest}$ configuration of antennas based on the evaluation of the fitness function 
\begin{equation}
    f^{(i)}_m = p^\mathrm{tx}_m + \tau |\mathcal{D}_m|,  
\end{equation}
where $p^\mathrm{tx}_m$ is the transmit power, $\mathcal{D}_m$ is a set containing the events of inter-antenna spacing violations, \textit{i.e.}, 
\begin{equation}
    \mathcal{D}_m = \{(\mathbf{r}_{n,m}, \mathbf{r}_{n',m}) |\ \lVert \mathbf{r}_{n,m} - \mathbf{r}_{n',m} \rVert_2 < \delta, n,n' \in \mathcal{N}, n < n' \},
\end{equation}
and $\tau$ is a large parameter utilized to penalize solutions that violate \eqref{P1d}. The transmit power $p^\mathrm{tx}_m$ required for particle $\mathbf{Q}_m$ can be obtained by first finding a unit-power analog precoder $\mathbf{w}_m$ that maximizes the minimum devices' received power, and then scaling $p^\mathrm{tx}_m$ to meet the constraint \eqref{P1c}. We can cast this problem into the following SDP formulation  
\vspace{-0.2em}
\begin{subequations}\label{P2}
    \begin{alignat}{2}
    \mathbf{P2:} \quad &\underset{\mathbf{W}_m,\ \xi}{\mathrm{max.}} \ && \xi \label{P2a}\\
    &\text{s.t.} \ && \mathrm{tr}(\mathbf{H}_{k,m}\mathbf{W}_m) \geq \xi , \ \forall k \in \cal K, \label{P2b}\\
    & \ && \mathbf{W}_m \succeq 0, \label{P2c}
    \end{alignat}
\end{subequations}
\vspace{-0.2em}
where $\mathbf{H}_{k,m} = \mathbf{h}_{k,m} \mathbf{h}^H_{k,m}$, $\mathbf{W}_m = \mathbf{w}_m\mathbf{w}^H_m$, and $\xi$ is an auxiliary variable in the epigraph formulation. We have relaxed the rank$-1$ requirement of $\mathbf{W}_m$ in \eqref{P2c} to make the problem convex. Once $\mathbf{P2}$ is solved, one can recover a suboptimal solution via Gaussian randomization, \textit{i.e.}, we generate multiple analog precoders $\mathbf{\hat{w}}_m$ whose phases are extracted from complex vectors with distribution $\mathcal{CN}(\mathbf{0}, \mathbf{W}_m)$, and then compute
\begin{equation}\label{eq:optPowAllocation}
    p^\mathrm{tx}_m = \underset{\mathbf{\hat{w}}_m}{\mathrm{min}}\ \underset{k}{\mathrm{max}} \ \frac{p^\mathrm{th}_k}{|\mathbf{h}_{k,m} \mathbf{\hat{w}}_m|^2},
\end{equation}
which is the allocated power to the precoder that maximizes the minimum received power among all devices, according to $\mathbf{P2}$. Notice that for the single-device scenario, analog maximum ratio transmission becomes optimal and hence the transmit power is $p^\mathrm{tx}_m = \frac{p^\mathrm{th}}{\lVert \mathbf{h}_m \rVert^2_1}$.

At the $i^\mathrm{th}$ iteration, the SgPSO algorithm updates the velocity and position of each particle as
\begin{subequations}
    \begin{align}
        \mathbf{V}^{(i)}_m &= \omega^{(i)}\mathbf{V}^{(i-1)}_m 
        + c_1 \mathbf{E}_1 \odot (\mathbf{Q}^{(i-1)}_{\mathrm{pbest},m} - \mathbf{Q}^{(i-1)}_m) \nonumber \\
        &\quad + c_2 \mathbf{E}_2 \odot (\mathbf{Q}^{(i-1)}_\mathrm{gbest} - \mathbf{Q}^{(i-1)}_m), \label{eq:velocityUpdate}\\
        \mathbf{Q}^{(i)}_m &= P_\mathcal{T}\left(\mathbf{Q}^{(i-1)}_m + \mathbf{V}^{(i)}_m\right), \label{eq:positionsUpdate}
    \end{align}
\end{subequations}
where $\odot$ is the element-wise matrix multiplication and $\omega^{(i)} = \omega_\mathrm{max} - (\omega_\mathrm{max} - \omega_\mathrm{min})i/I$ is the inertia weight of each particle's velocity with $I$ is the total number of iterations. Moreover, $c_1$ and $c_2$ are the personal and global learning factors, respectively, whereas $\mathbf{E}_1$ and $\mathbf{E}_2$ are two random matrices whose components are uniformly distributed in the interval $[0,\ 1]$. These factors aid the SgPSO algorithm in balancing the exploration and exploitation phases, preventing the algorithm from getting trapped in a local optimum in early iterations. $P_\mathcal{T}(\cdot)$ confines the $x$ and $y$ coordinates of any antenna position $\mathbf{r}$ onto the set $\mathcal{T}$, \textit{i.e.},
\begin{equation}
    P_\mathcal{T}(r_x) = 
        \begin{cases}
            \frac{l}{2}, & \text{if} \quad r_x > \frac{l}{2}, \\
            r_x, & \text{if} \quad -\frac{l}{2} \leq r_x \leq \frac{l}{2}, \\
            -\frac{l}{2}, & \text{if} \quad r_x < -\frac{l}{2}.
        \end{cases} \label{eq:projection}
\end{equation}
Moreover, the SgPSO algorithm may update each particle's personal best and the swarm's global best as
\begin{subequations}
    \begin{align}
        \mathbf{Q}^{(i)}_{\mathrm{pbest},m} &= 
    \begin{cases}
        \mathbf{Q}^{(i)}_m, & \text{if} \quad f^{(i)}_m < f_{\mathrm{pbest},m}, \\
        \mathbf{Q}^{(i-1)}_{\mathrm{pbest},m},               & \text{otherwise},   
    \end{cases} \label{eq:personalBestUpdate}\\
    \mathbf{Q}^{(i)}_{\mathrm{gbest}} &= 
    \begin{cases}
        \mathbf{Q}^{(i)}_{m^*}, & \text{if} \quad \underset{m}{\mathrm{min}} \ f^{(i)}_m < f_{\mathrm{gbest},m}, \\
        \mathbf{Q}^{(i-1)}_{\mathrm{gbest}},               & \text{otherwise},   
    \end{cases} \label{eq:swarmGlobalUpdate}
    \end{align}
\end{subequations}
where $m^* = \underset{m}{\mathrm{argmin}} \ f^{(i)}_m$.

\begin{algorithm}[t!]
\caption{SDP-guided PSO (SgPSO)}
\begin{algorithmic}[1] \label{alg:PSO}
\STATE \textbf{Input:} $\{\mathbf{u}_k\}, \{p^\mathrm{th}_k\}, N, M$ \label{alg:PSO1}
\STATE Randomly initialize ${\mathbf{Q}_m}$ and ${\mathbf{V}_m}$ \label{alg:PSO2}
\WHILE{$i \leq I$} \label{alg:PSO3}
\FOR{$\mathbf{Q}_m \in \mathcal{Q} $} \label{alg:PSO4}
    \STATE Update $\mathbf{V}^{(i)}_m$ using \eqref{eq:velocityUpdate} \label{alg:PSO5} 
    \STATE Update $\mathbf{Q}^{(i)}_m$ using \eqref{eq:positionsUpdate} \label{alg:PSO6}
    \STATE Evaluate fitness function by solving $\mathbf{P2}$ \label{alg:PSO7}
    \STATE Update particle's personal best using \eqref{eq:personalBestUpdate} \label{alg:PSO8}
    \ENDFOR \label{alg:PSO9}
    \STATE Update swarm's global best using \eqref{eq:swarmGlobalUpdate} \label{alg:PSO10}
    \STATE Update the inertia weight \label{alg:PSO11}
    \STATE $i \gets i + 1$ \label{alg:PSO12}
\ENDWHILE \label{alg:PSO13}
\STATE \textbf{Output:} $\{\mathbf{r}^*_n\}$, $\mathbf{w^*}$ \label{alg:PSO14}
\end{algorithmic}
\end{algorithm}

At each iteration, SgPSO requires solving $M$ SDPs. Interior point methods (IPMs) find an $\epsilon-$optimal solution of an SDP in $\mathcal{O}(\sqrt{N}\log{\frac{1}{\epsilon}})$ iterations, with $\mathcal{O}(\cdot)$ being the worst-case complexity, each requiring $(K+1)(N+1)^2 + (K+1)^2(N+1)^2 + (K+1)^3$ arithmetic operations, thus costing $\mathcal{O}\left(K^2N^2 + K^3\right)$ in the worst case \cite[Theorem 3.12]{Bomze.2010}. Overall, the objective function evaluation, which is the most critical step in the SgPSO algorithm, requires $\mathcal{O}(IM\sqrt{N}\log{\frac{1}{\epsilon}})$ iterations and $\mathcal{O}\left(IMK^2N^2 + IMK^3\right)$ arithmetic operations per iteration. Hence, the computational complexity is dominated by the number of devices when $K > N^2$, and by the number of antennas otherwise. This dependency on complexity in the number of particles, antennas, and devices can make SgPSO relatively slow in the massive antennas/devices regime. Next, we show an alternative architecture consisting of UMAs with homogeneous separation between adjacent elements.

\vspace{-1.2em}
\section{SgSDP for UMAs}\label{sec:l1MaxMinHeuristic}
Optimizing the configuration of UMAs can reduce the computational complexity as only the reference position of the array, its rotation, and the separation between adjacent elements need to be determined.
\begin{figure}[h]
    \centering
    \includegraphics[width=\linewidth]{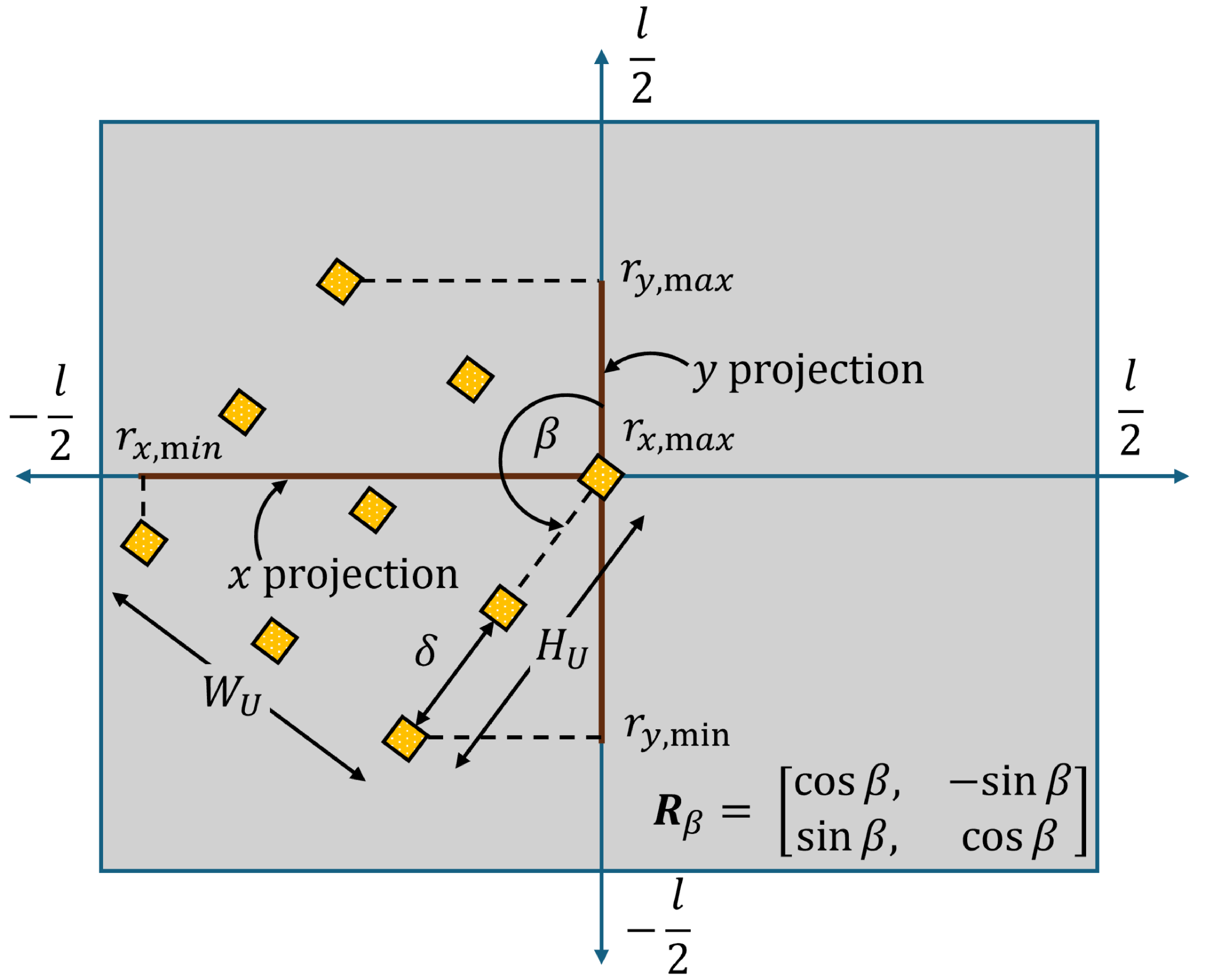}
    \vspace{-2em}
    \caption{Geometry of the UMAs rotated counterclockwise by an angle $\beta$.}
    \label{fig:UMAsGeometry}
\end{figure}
As illustrated in Fig.~\ref{fig:UMAsGeometry}, the antenna elements are arranged on a grid resembling a uniform rectangular array (URA), with $N_x = \lceil \sqrt{N} \rceil$ columns and $N_y = \lceil N/N_x \rceil$ rows, \textit{i.e}, with height $H_\mathrm{U} = (N_y - 1)\delta_U$ and width $W_\mathrm{U} = (N_x - 1)\delta_U$, where $\delta_U$ is the inter-antenna separation. The resulting grid may contain more positions than antennas, i.e., $N_x N_y > N$, and thus the array is not necessarily fully populated. Then, considering $\mathbf{r}_0$ as the reference position of the array, the position of the $n^\mathrm{th}$ antenna is
\begin{equation}
    \mathbf{r}_n = \mathbf{r}_0 + R_\beta\begin{bmatrix} \lfloor n/N_x \rfloor \delta & n\ \mathrm{mod}\ N_x \end{bmatrix}^T,
\end{equation}
where $\mathbf{R}_\beta \in \mathbb{R}^{2 \times 2}$ is a rotation matrix in the $xy$ plane counterclockwise by $\beta$ radians. SgSDP can be reused by redefining the $m^\mathrm{th}$ particle as $\mathbf{Q}_m = [\mathbf{r}_{0,m}^T,\ \beta_m,\ \delta_{U,m}]^T$,
containing the reference position, orientation, and inter-antennas separation, with velocity $\mathbf{V}_m \in \mathbb{R}^{4 \times 1}$. Moreover, $P_\mathcal{T}(\cdot)$ confines $\delta_U$ to the interval $[\lambda/2,\ \delta_{U,\mathrm{max}}]$, where
\begin{equation}
\delta_{U,\mathrm{max}} = l \left\lVert\
\begin{bmatrix}
    |\cos{\beta}|, & |\sin{\beta}| \\
    |\sin{\beta}|, & |\cos{\beta}|
\end{bmatrix}
\begin{bmatrix}
N_{x}-1\\
N_{y}-1
\end{bmatrix}
\right\rVert^{-1}_{\infty}
\end{equation}
is the maximum inter-antennas separation which is obtained by restricting the $x$ and $y$ projections of the UMAs to be lower than $l$. Moreover, the $x$ coordinate of the reference position $\mathbf{r_0}$ is confined to $[-l/2 - r_{x,\mathrm{min}},\ l/2 - r_{x,\mathrm{max}}]$, where
\begin{subequations}
    \begin{align}
        r_{x,\mathrm{min}} &= \underset{\{r_x, r_y\} \in \{0,W_\mathrm{U}\} \times \{0,H_\mathrm{U}\}}{\mathrm{min}} \begin{bmatrix}
            1 \\ 0
        \end{bmatrix}^T\mathbf{R}_\beta \begin{bmatrix}
            r_x \\ r_y
        \end{bmatrix},\\
        r_{x,\mathrm{max}} &= \underset{\{r_x, r_y\} \in \{0,W_\mathrm{U}\} \times \{0,H_\mathrm{U}\}}{\mathrm{max}} \begin{bmatrix}
            1 \\ 0
        \end{bmatrix}^T\mathbf{R}_\beta \begin{bmatrix}
            r_x \\ r_y
        \end{bmatrix}.
    \end{align}
\end{subequations}
Similar restrictions are applied on the $y$ coordinates and only the basis vector changes to $[0,\ 1]$. The SgPSO algorithm has the same worst-case complexity form derived in Section~\ref{sec:PSOSolution} as $\mathbf{P2}$ still requires to be solved. However, since each particle is now a four-element decision vector, the search space no longer expands with $N$. Consequently, SgPSO requires fewer particles and iterations to reach convergence.
\section{Numerical results}\label{sec:numericalResults}
Unless otherwise specified, we consider three devices, whose power requirement is $p^\mathrm{th}_k = 1$~mW, uniformly distributed in a plane with dimensions $a_x \times a_y$ located at a distance $a_z$ from the IMAs, as shown in Fig.~\ref{fig:systemModel}. By default, we set $a_x = a_y = 8~$m and $a_z = 3~$m, covering a zone within a factory facility. Moreover, we set $\delta = \lambda/2$ where $\lambda$ is the wavelength of a $1~$GHz signal. The inertia weight parameter varies within the range $[\omega_\mathrm{min}, \omega_\mathrm{max}] = [0.1, 1]$, while the penalty parameter is set to $\tau = 10^4$. The personal and global learning factors are both set to $1.49$. The SgPSO runs for $I = 200S$ iterations, utilizing $M = \mathrm{min}(150,10S)$ particles, where $S$ equals $2N$ and $4$ for the IMAs and UMAs architectures, respectively. As a benchmark, we consider a system where the PB is equipped with uniform linear (ULA) or URA arrays with fixed elements centered at the origin, arranged along the $x-$axis or within the $x-y$ plane, respectively. For them, adjacent elements are separated by $\delta$, and power allocation is determined by solving $\mathbf{P2}$. A suboptimal solution of $\mathbf{P2}$ is obtained using $10^6$ candidate vectors $\Tilde{\mathbf{w}}_m$ in all cases. All results are averaged over $100$ deployments.

\begin{figure}
    \centering
    \includegraphics[width=\linewidth]{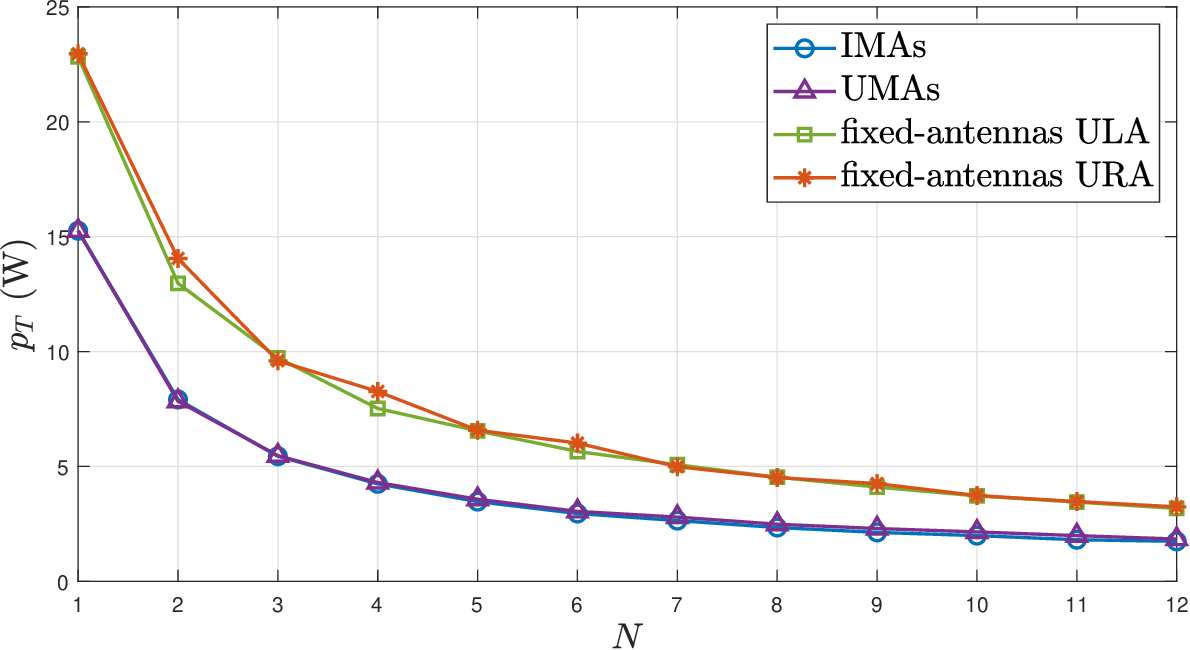}
    \vspace{-1.3em}
    \caption{PB's power consumption vs $N$ for $K=3$ and $l=1~$m.}
    \label{fig:powVSN}
\end{figure}

Fig.~\ref{fig:powVSN} shows the PB's power consumption against the number of transmitting antennas for a fixed number of deployed devices. The power consumption decreases as the number of antennas increases for all architectures. Observe that the IMA-equipped PB outperforms the proposed benchmark architectures due to its capacity to efficiently distribute the antennas to meet individual devices' power demands. However, the gap with respect to the UMAs architecture is negligible thereby indicating the effectiveness of the UMAs, which have lower hardware/configuration complexity. Additionally, equipping the PB with a URA does not provide gains over the ULA case, as the aperture of the latter grows faster with $N$.  

\begin{figure}
    \centering
    \includegraphics[width=\linewidth]{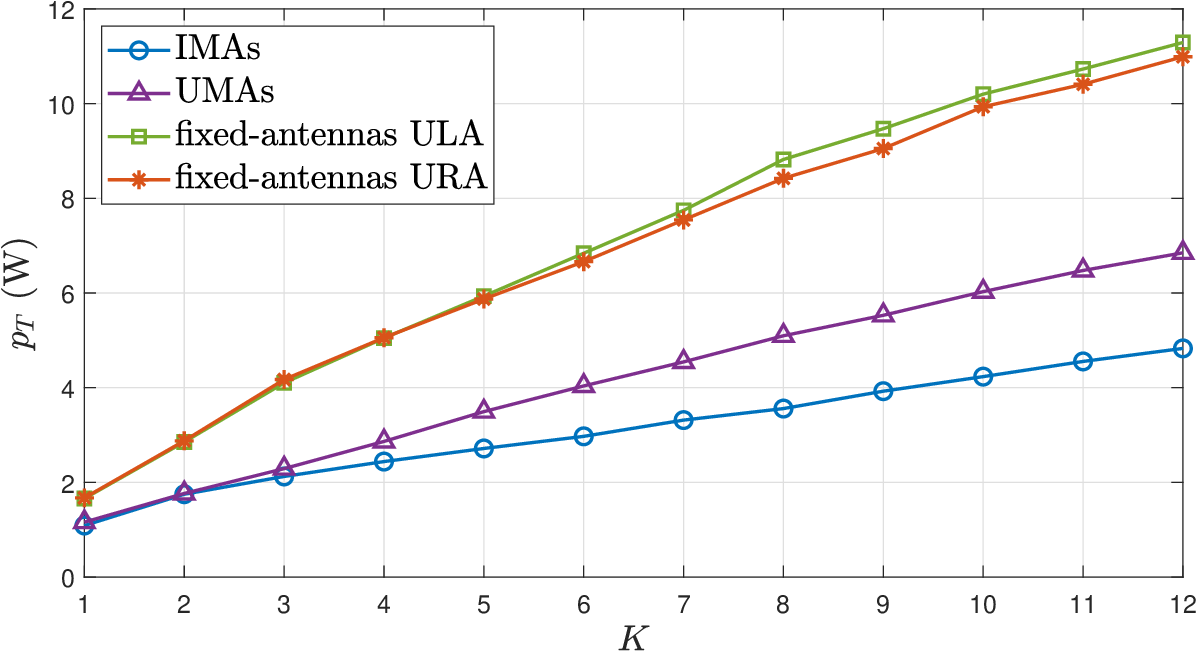}
    \vspace{-1.3em}
    \caption{PB's power consumption vs $K$ for $N=9$ and $l=1~$m.}
    \label{fig:powVSK}
\end{figure}

Fig.~\ref{fig:powVSK} illustrates how the power consumption increases with the number of deployed devices for all architectures. The architecture with IMAs outperforms its counterparts with fixed antennas and the UMAs, with an increasing performance gap as the number of devices increases. Observe that the achieved solution with the URA slightly outperforms that employing a ULA, as the URA's two-dimensional beam-steering capability allows it to better align its analog beam to cover a large number of spatially distributed devices, thereby reducing required transmit power. The UMAs outperforms the arrays with fixed antennas due to its reconfiguration to meet devices power needs. Notably, its limited degrees-of-freedom with respect to the IMAs penalize its performance as $K$ grows.

\begin{figure}
    \centering
    \includegraphics[width=\linewidth]{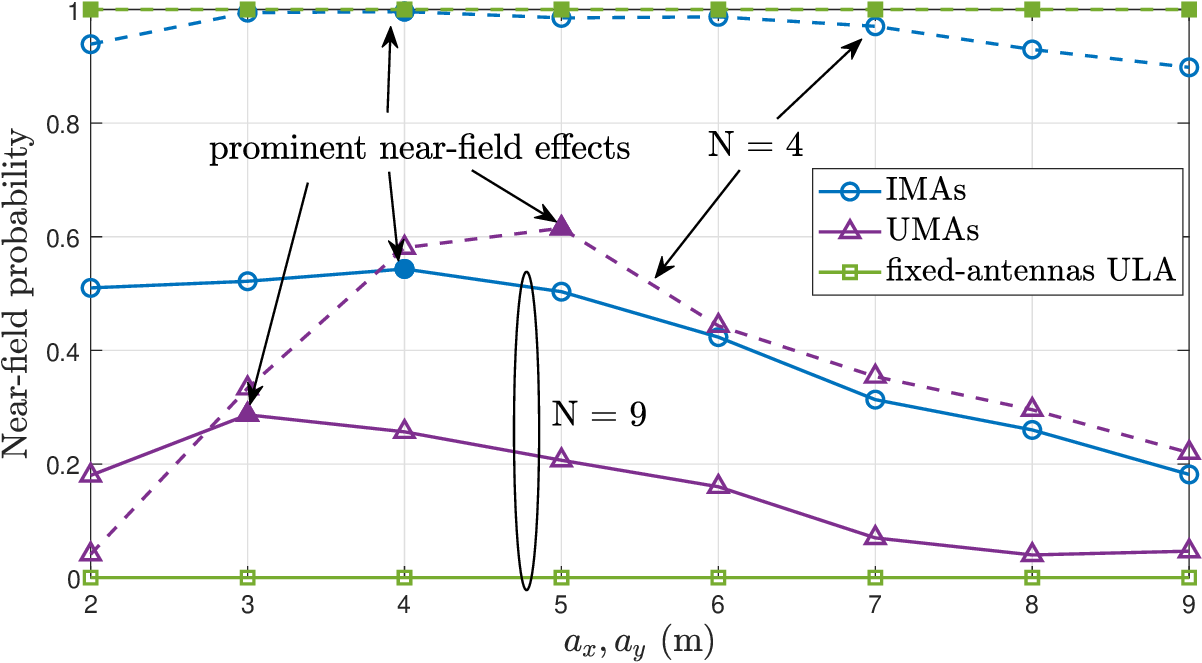}
    \vspace{-1.3em}
    \caption{Probability of operation in the near-field for varying distributions of the network deployment for $K = 3$ and $l = 1~$m. Since $2D^2/\lambda = 1.2$~m in the URA with $N = 9$, its probability of near-field operation is zero.}
    \label{fig:nfProb}
\end{figure}

Finally, we examine how variations in the antenna aperture and the distribution of devices across network with different sizes, \textit{i.e.}, $a_x \times a_y$, influence channel propagation conditions. We are interested in the likelihood of having the devices in the near-field region for different values of $a_x = a_y$. To this end, we establish that a device operates in the near-field if $\left\lVert \begin{bmatrix} \mathbf{r}_c, & 0 \end{bmatrix}^T - \mathbf{u}_k \right\rVert_2 \leq \frac{2D^2}{\lambda}$, where $\mathbf{r}_c \in \mathbb{R}^2$ and $D$ are the centroid and aperture---\textit{i.e.}, the largest dimension of the convex hull comprising the set of antennas---of the array, respectively \cite{Kosasih.2024}. These results are shown in Fig.~\ref{fig:nfProb} for different number of transmit antennas. Notice that increasing the number of antennas promotes the operation in the near-field. For the ULA case, the probability is either $0$ or $1$ as the aperture remains constant for a given $N$. When devices are closely clustered in small areas, the optimal placement of the MAs tends to be concentrated in a limited region. Conversely, as we expand the service area limits, the probability of having near-field conditions decreases despite the antennas being more widely spaced on average. This occurs because the devices can be positioned further away from the convex hull of the array, thereby increasing the operation distance compared to the effective aperture of the array. This effect becomes more prominent for the IMAs architecture than for the UMAs, as in the former the antennas move independently. The filled markers indicate the points where there is more control over energy beaming and potentially on electromagnetic field radiation, although not necessarily with the lowest power consumption.

\section{Conclusions}\label{sec:conclusions}
We investigated the power consumption of an MA-equipped PB for charging an IoT network. We formulated a nonconvex optimization problem considering device power requirements, minimum separation between adjacent antennas to avoid coupling, and the corresponding constraints on the analog precoder. To overcome the complexity of the resulting problem, we proposed the SgPSO algorithm to explore different antenna positions while evaluating the effectiveness of each position by solving the corresponding power allocation problem. We also evaluated an architecture with uniform inter-antenna spacing whose SgPSO requirements of particles and iterations do not scale with the number of antennas. Results evinced the effectiveness of the MAs in reducing power consumption, with UMAs performing close to the IMAs; only the gap increases when serving a large number of devices simultaneously 
due to its limited degrees of freedom with respect to the IMAs. Finally, we showed that increasing the number of antennas enhances near-field operation, while expanding the service area beyond certain limits reduces the near-field operation probability for the MAs due to increased network dispersion relative to the array's aperture.

\bibliographystyle{IEEEtran}
\bibliography{IEEEabrv,referencesShort}
\end{document}